\documentclass[%
preprint,
amsmath,amssymb,
aps,
floatfix,
superscriptaddress,
]{revtex4-1}
\usepackage{graphicx}
\usepackage{dcolumn}
\usepackage{bm}
\usepackage{placeins} 
\usepackage{miller}
\usepackage{siunitx} 
\usepackage{tikz}
\usepackage{comment}
\usepackage{graphicx}
\usepackage[T1]{fontenc} 

\usepackage[unicode=true, 
 bookmarks=true,bookmarksnumbered=false,bookmarksopen=false,
 breaklinks=true,pdfborder={0 0 1},backref=false,colorlinks=true]
 {hyperref}
\hypersetup{pdftitle={Mapping of Local Lattice Parameter Ratios by Projective Kikuchi Pattern Matching},
 pdfauthor={Aimo Winkelmann}}

\begin{document}

\title{Mapping of Local Lattice Parameter Ratios by Projective Kikuchi Pattern Matching}

\author{Aimo Winkelmann}
\email{a.winkelmann@lzh.de}
\affiliation{Laser Zentrum Hannover e.V., Hollerithallee 8, 30419 Hannover, Germany }

\author{Gert Nolze}
\affiliation{Federal Institute for Materials, Research and Testing (BAM), Unter den Eichen 87, 12205 Berlin, Germany}
\email{gert.nolze@bam.de}

\author{Grzegorz Cios}
\affiliation{Academic Centre for Materials and Nanotechnology, AGH University of Science and Technology, al.\@ A. Mickiewicza 30, 30-059 Krakow, Poland}
\email{ciosu@agh.edu.pl}

\author{Tomasz Tokarski}
\email{tokarski@agh.edu.pl}
\affiliation{Academic Centre for Materials and Nanotechnology, AGH University of Science and Technology, al.\@ A. Mickiewicza 30, 30-059 Krakow, Poland}

\date{\today}

\begin{abstract}
We describe a lattice-based crystallographic approximation for the analysis of distorted crystal structures via Electron Backscatter Diffraction (EBSD) in the scanning electron microscope. 
EBSD patterns are closely linked to local lattice parameter ratios via Kikuchi bands that indicate geometrical lattice plane projections.
Based on the transformation properties of points and lines in the real projective plane, we can obtain continuous estimations of the local lattice distortion based on projectively transformed Kikuchi diffraction simulations for a reference structure. 
By quantitative image matching to a projective transformation model of the lattice distortion in the full solid angle of possible scattering directions, we enforce a crystallographically consistent approximation in the fitting procedure of distorted simulations to the experimentally observed diffraction patterns.
As an application example, we map the locally varying tetragonality in martensite grains of steel.
\end{abstract}

\maketitle


The characterization of locally changing lattice parameter ratios within crystalline materials is crucial for an improved understanding of the properties of technological and geological materials in the presence of varying mechanical, thermal, or chemical conditions, which can induce crystal structure distortions on microscopic scales. 
Here, we present an approach to map the lattice parameter ratios of a material with respect to continuous perturbations from a reference structure. This is made possible by application of the method of electron backscatter diffraction (EBSD) for sub-micron scale crystallographic mapping in the scanning electron microscope (SEM).

EBSD delivers spatially resolved crystallographic information via the measurement of Kikuchi patterns that are formed by incoherent point sources within a crystal structure \cite{schwartzEBSD,venables1973pm}.
The observation of shifting Kikuchi pattern features on a geometrically calibrated phosphor screen provides the possibility to measure small local distortions of a material \cite{troost1993apl}, with resolutions in the order of  $10^{-4}$ \cite{wilkinson2006um}.
While the achievable precision of high-resolution EBSD (HR-EBSD) measurements is very attractive for a wide range of applications, these experiments also involve a substantially increased complexity in the experimental setup, the data acquisition, and the data analysis.
For example, advanced Kikuchi pattern simulations \cite{winkelmann2007um,callahan2013mm} would be highly useful for a comparison to absolutely known reference structures via HR-EBSD, but the application of such patterns is limited by the simulation speed when using dynamical diffraction theory. 
This is why it could often be desirable to (a) have an initial option for controlled compromises in accuracy and precision when trying to map continuous changes of crystal lattices, e.g. via routine EBSD measurements, and then (b) being able to subsequently refine these results using iteratively improving EBSD data analysis approaches.

With these demands in mind, the example application which we discuss in this paper concerns the problem of martensite tetragonality in carbon containing steels \cite{bhadeshia2011steels}. 
With the average lattice parameters depending on the carbon content \cite{cheng1990smet}, a tetragonally distorted martensite forms after rapid quenching of the high temperature face-centered cubic austenite phase. 
To characterize the martensitic microstructure in more detail, it would be beneficial to monitor the local tetragonality with microscopic resolution and compare the results to X-ray diffraction measurements that lack the spatial resolution of EBSD.

The locally resolved investigation of martensite tetragonality has been approached previously by comparison to kinematic simulations of Kikuchi patterns \cite{kacher2009thesis}. Also, dynamical electron diffraction simulations for a fixed number of lattice parameters have been combined with subsequent pattern cross-correlation \cite{tanaka2018mm}. Moreover, a similar study of tetragonally distorted SiGe films has been published in \cite{jackson2016mm}.

In the current paper, we describe an approach which is tailored to the different sensitivity of specific diffraction pattern features under a distortion of the crystal lattice. 
We will demonstrate that the underlying geometrical and physical conditions of Kikuchi diffraction pattern formation allow us to approximate the effects of the local distortion of the crystal lattice by the resulting \textit{projected} effects on reference diffraction data.
This makes it possible to estimate the local lattice distortion in a sample by pattern matching to a single global simulation for a particular reference structure, which is stored for the full $4\pi$ solid angle of possible scattering directions. No additional electron diffraction simulations for the perturbed test lattices are required during the fit of the local lattice distortion. 
Because the projective crystallographic approximation improves in the limit of small changes from the reference lattice, results of our approach can be iteratively refined.

\begin{figure}[htbp]  %
	\includegraphics[width=0.85\textwidth]{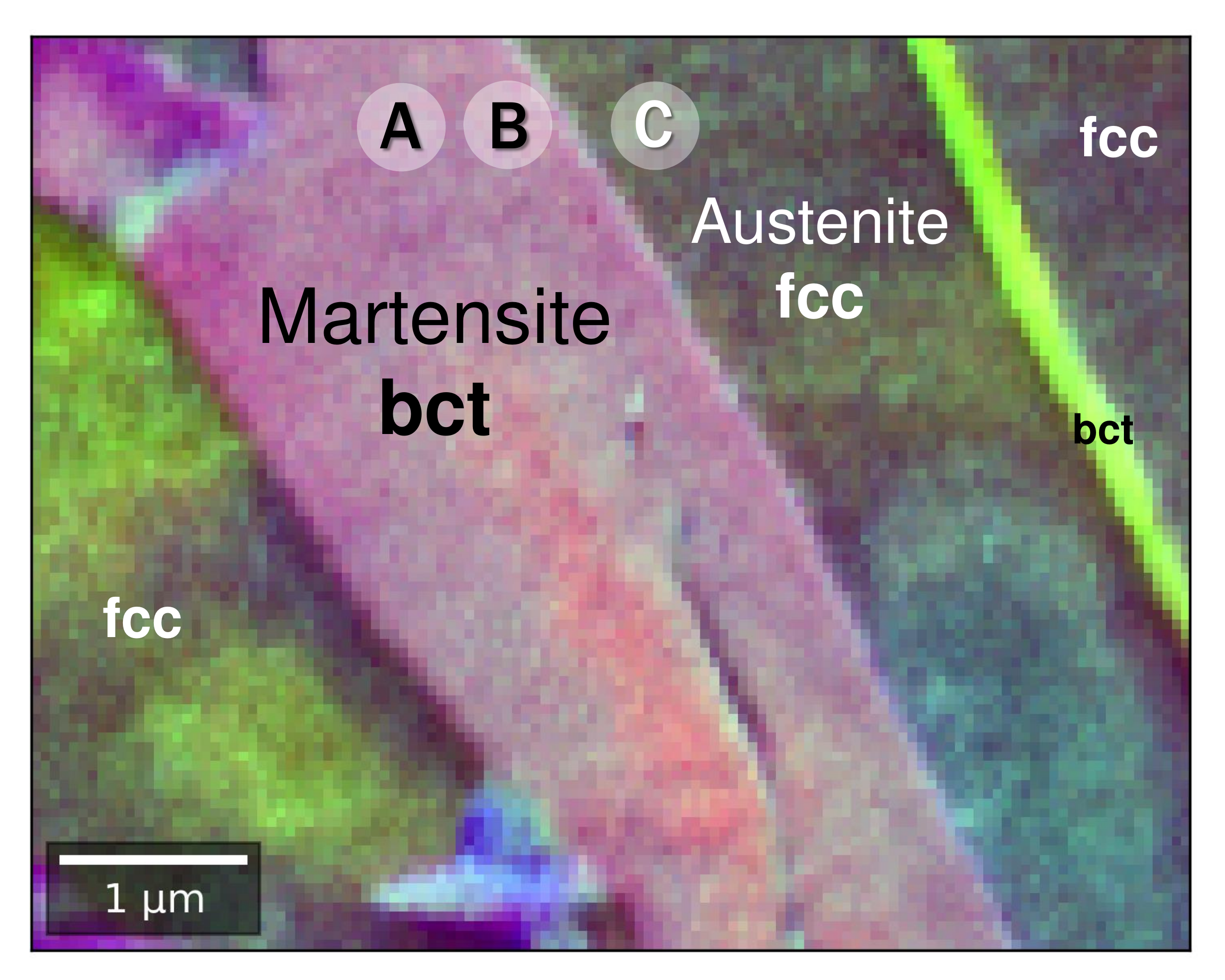}
	\caption{Pseudocolor SEM image derived from the EBSD Kikuchi patterns of the measured sample area, with indicated areas of martensite (bct) and austenite (fcc). The labels A, B, and C mark reference positions used for the discussion of the pattern analysis as shown in Fig.\,\ref{fig:bct_lattice} and Fig.\,\ref{fig:patterns}.}
	\label{fig:sem}
\end{figure}

\textit{Sample Description.}
For our investigation, we produced a steel sample, for which we confirmed from X-ray investigations an average $c/a$-ratio for the martensite of about $c/a=1.07$ (for details for the X-ray equipment see \cite{cios2017mmta}).
In Fig.\,\ref{fig:sem}, we show a pseudocolor SEM image of the measured sample region, where the body-centered tetragonal (bct) martensite and face-centered cubic (fcc) austenite grains are indicated.
The image in Fig.\,\ref{fig:sem} has been derived from the measured EBSD patterns \cite{wright2015um,brodusch2015um,winkelmann2017jm} by assigning color variations to variations in intensity between regions of interest (ROI) in the resulting background-processed Kikuchi patterns. 
Specifically, we have binned each processed Kikuchi pattern to a $7\times 7$ ROI array and calculated the ratio of the ROIs in column 3,4,5 between row 6 and row 5, respectively. The variation in the ROI intensity ratio has been mapped to the range of the R,G,B color channels of the image.
This ``differential Kikuchi pattern imaging'' mode qualitatively emphasizes  orientation changes between and within the different grains, which are seen by color changes in the image. 

\textit{Method.}
Kikuchi pattern features which are observed on a planar phosphor screen are closely related to the gnomonic projection of crystallographic features like lattice planes \hkl(hkl) and lattice directions \hkl[uvw], which appear as line traces and intersections of these linear traces, respectively \cite{nolze2017jac}. 
Changes in these geometrically defined features of Kikuchi patterns can thus be variably linked to changes in the projection geometry, to lattice rotations, but also to variations in the crystal lattice parameters.
In current HR-EBSD approaches \cite{wilkinson2012mt} and suggested improvements related to general feature registration \cite{maurice2014rms,ruggles2018um,vermeij2018um}, the specifically different crystallographic nature of the various observed pattern features remains largely untapped. This means that the Kikuchi features mostly serve as reference points in a coordinate system in which the deformation of the material can be measured via feature matching like in digital image correlation (DIC) methods \cite{sutton2009dic}. The basic feature-shift approach of HR-EBSD could work even with non-crystallographic types of angle-dependent emission patterns as long as one can reliably identify spatial shifts of features on a geometrically calibrated phosphor screen.

With respect to the measurement of lattice directions and lattice plane normals, crystal geometry \cite{miller1859pm,liebisch1881geo,buerger1934am} is essentially an experimental realization of abstract mathematical \emph{projective geometry}, specifically of the real projective plane $\mathbf{RP}^2$ \cite{coxeter1987projective_geometry}.
Backscatter Kikuchi Diffraction (BKD) patterns measured at typical tens of keV energies using EBSD in the SEM represent a fortunate case where electron scattering physics emphasizes the relation of mathematical projective geometry to the geometry of the crystal lattice.
The predominantly forward peaked electron scattering in the diffraction process limits the scattered electron intensity to the lowest reflection orders of a lattice plane (an instructive counterexample is given by neutron $s$-wave scattering which leads to very different patterns for an equivalent crystal geometry \cite{winkelmann2017crat}).
Because electron Kikuchi band widths are relatively small ($\lessapprox5^\circ$ for the strongest bands) and change only slowly with changes in crystal structure or scattering energy, these physical bands can outline the mathematical traces of the relevant lattice planes in a relatively stable way.

\begin{figure}[t!]  
    \vspace{-1cm}
\includegraphics[width=10.5cm]{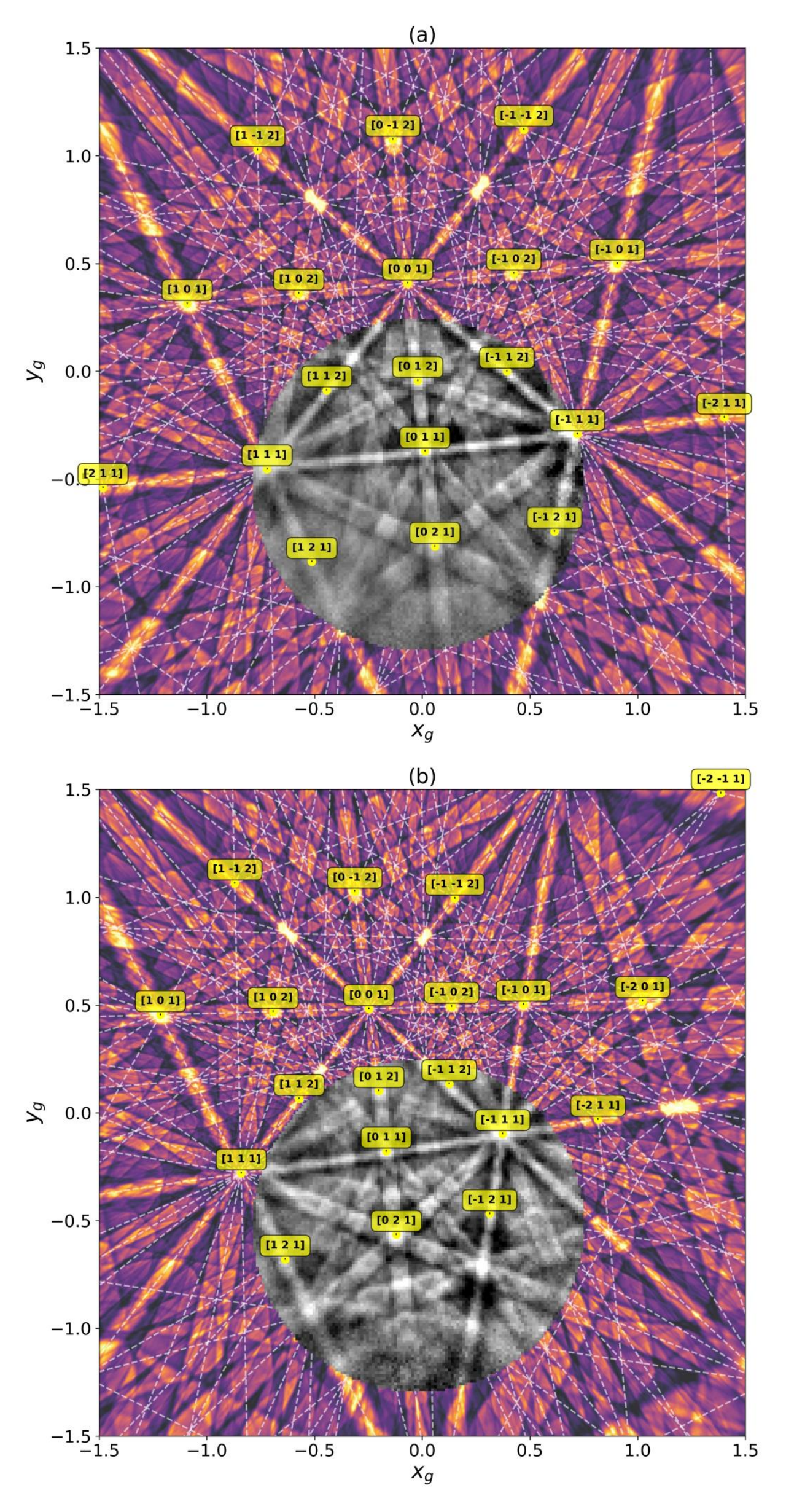}
	\caption{Projective Kikuchi pattern transformation.  
	(a): experimental \textit{bct} Kikuchi pattern (gray) from position A in Fig.\ref{fig:sem}, and dynamical simulation (color) according to $c/a=1.08$\\
	(b): experimental \emph{fcc} pattern (gray) from position C, with the \emph{same} dynamical \emph{bct} simulation data as in (a) \emph{projectively  transformed} according to a resulting $c/a=1.41$.
	(See text for explanation of the axes ranges)}
	\label{fig:bct_lattice}
\end{figure}

In order to make the connection between the experimental \emph{projection geometry} for Kikuchi pattern measurements and its mathematical description via concepts of \emph{projective geometry}, in Fig.\,\ref{fig:bct_lattice}\,(a) we display experimental BKD patterns (gray) collected at sample positions "A"  and "C" in Fig.\,\ref{fig:sem}.
At location A, we find bct-Fe with $c/a=1.08$ as the best fit to a simulation which is shown as the pattern in color in the background.
For comparison, the experimental pattern at location C corresponds to fcc-Fe.
The experimental patterns have been calibrated in a 2D plane of coordinates $(x_g, y_g, z_g)=(x_g, y_g, 1)$ in such a way that the 2D distance between the point \hkl[uvw] and $[x_g, y_g, z_g]=[0,0,1]$ corresponds to $\tan\theta$ of the angle $\theta$ between the crystal direction \hkl[uvw] and the $z$ axis direction.
Regarding the Kikuchi patterns in Fig.\,\ref{fig:bct_lattice}, this calibration means that the pattern center will always be at (0,0) and the coordinates of the four corner points of the measured pattern define the extent in the projection plane.
While this is commonly known as a gnomonic projection, the plane $(x_g, y_g, 1)$ also corresponds to the standard embedded Euclidean plane in analytic projective geometry using homogeneous coordinates. In this way, crystal directions correspond to points, while lattice plane traces correspond to lines \cite{miller1859pm,liebisch1881geo,buerger1934am}. 
Via this correspondence, knowledge about the real projective plane $\mathbf{RP}^2$ can be applied to the crystallographic analysis of experimental \mbox{Kikuchi patterns \cite{alkorta2017um}}. 

For example, the image analysis pipeline \cite{britton2016mc} which is pertinent to the gnomonic projection of lattice directions in Kikuchi diffraction patterns \cite{nolze2017jac} can be understood in a unified way by concepts of projective geometry, with a projective transformation $\mathbf{H}$ acting on the homogeneous coordinates \hkl[uvw] \cite{alkorta2017um}:
\begin{equation} 
    \begin{bmatrix}
        p_1 \\
        p_2 \\
        p_3  
    \end{bmatrix} = \mathbf{H} \begin{bmatrix}
        u \\
        v  \\
        w  
    \end{bmatrix}
    = \mathbf{P}\,\mathbf{O}\,\mathbf{C} \begin{bmatrix}
        u \\
        v  \\
        w  
    \end{bmatrix}
\end{equation}

The coordinates $(x_g, y_g)$ in the standard embedded projection plane at $z_g=1$ are given by dehomogenization of the projective coordinates as $(x_g, y_g) = (p_1/p_3,\, p_2/p_3)$ for all projected points with $p_3 \neq 0$. $\mathbf{H}$ has 8 degrees of freedom, so that scaling of $\mathbf{H}$ by a real factor does not change the projective transformation \cite{hartley2004MVG}. 
The homogeneity of coordinates \hkl[uvw] is also consistent with the fact that the present HR-EBSD approaches are not directly sensitive to the hydrostatic expansion of the lattice \cite{villert2009jm}.

For the interpretation of $\mathbf{H}$ in terms of the orientations $\mathbf{O}$ which are measured by EBSD, the general $(3 \times 3)$ homography matrix $\mathbf{H}$ can be decomposed into the combined contributions of a transformation $\mathbf{C}$ from crystal coordinates to Cartesian coordinates, a rotation matrix $\mathbf{O}$,
 and the selection of the projected area on the screen relative to the projection center by the scaling and translation transformation $\mathbf{P}$  \cite{britton2016mc}.

The result of a tetragonal lattice distortion can be described by a corresponding transformation $\mathbf{F}$ of the reference matrix $\mathbf{C}$ to result in a transformed matrix $\mathbf{C_{T}}$ for a cubic reference:
\begin{equation} 
\mathbf{C_{T}} = \mathbf{F}\mathbf{C} =  \begin{pmatrix}
        1 & 0 & 0   \\
        0 & 1 & 0   \\
        0 & 0 & c/a  \\ 
    \end{pmatrix} \mathbf{C}
\end{equation}
with corresponding changes for arbitrarily strained lattices. Similar projective relations have been applied for EBSD strain analyses using kinematically simulated patterns \cite{jackson2016mm,alkorta2017um}.
With respect to dynamical pattern simulations we can in principle apply the same approach by noting that the tetragonality $c/a$ in the transformed matrix $\mathbf{C_{T}}$ changes the specific reference lattice direction \hkl[u_T v_T w_T] that is projected on the screen in the previous position of \hkl[u_0 v_0 w_0] of the non-distorted lattice. 
We can thus sample the diffracted intensity for the tetragonally distorted structure from the distorted global pattern data for the reference lattice. 
As viewed from the central point of the gnomonic projection inside a reference sphere of diffracted intensity, we essentially distort the sphere into an ellipsoid corresponding to the distortion of the lattice.
Because lattice zone axes \hkl[uvw] and the \hkl(hkl) lattice plane traces are transformed exactly, the high intensity in zone axes and in the center of Kikuchi bands will be transformed to positions which approximately correspond also to intensity maxima in the exact dynamical simulation for the distorted structure. 
Of course, this approximation requires that the crystal structure in the strained state retains sufficient similarity to the unstrained reference, in order that the diffraction patterns still show the corresponding Kikuchi bands with sufficient similarity. We can assume that this will be nearly always the case for small deviations from the reference structure, but this does not exclude rather large deviations.

This can be illustrated by returning to the experimental Kikuchi patterns shown in Fig.\,\ref{fig:bct_lattice}. We have used a projective transformation to transform the exactly simulated diffraction data for Fig.\,\ref{fig:bct_lattice}(a) into the approximate theoretical pattern shown in (b), e.g we apply a strain transformation to the global reference simulation which takes $c/a$ from $1.08$ to $1.41$ near the ideal $\sqrt{2}$ for the fcc structure. 
In the fit of the projectively transformed \textit{bct} global reference data to the \textit{fcc} pattern, we see that the dashed lines (lattice plane traces) and all their intersections (zone axes) are transformed \emph{exactly}, and thus the major experimental fcc bands still align approximately symmetric to the transformed bct-plane traces. The approximate alignment of major high-intensity Kikuchi bands centroids is the key reason why we still obtain sufficiently good values for the image similarity using the normalized cross correlation coefficient (see below).
In contrast, the transformed bct Kikuchi band \textit{widths} deviate significantly from the experimental fcc pattern, as can be seen by the discontinuities at the fcc pattern edge. As long as the deviations of the \textit{widths} are small, however, the good alignment of the band \textit{centroids} can dominate the fit.
Compared to the extreme case of a bcc-fcc transition, we will consider in the following only deviations of $c/a$ from the ideal bcc value of $c/a=1.0$ in the bct martensite itself. This will involve changes in $c/a$ ratios below 0.1 and only small rotations ($\pm 2^{\circ}$), which still can be handled in a stable way by the specific optimization approach which we apply in the pattern fit (see below).  
However, the example of Fig.\,\ref{fig:bct_lattice} shows that for the bcc-fcc structural relationship in Fe, the full ideal range of $c/a$-ratios can be handled by the approximative projective transformation \mbox{(\,$1_{(\textrm{bcc})} \leq c/a \leq \sqrt{2}_ {(\textrm{fcc})}$ ) } via the bct symmetry common to both structures.

We note that our approximations are purely geometrical, i.e. we make no assumptions about mechanical stress conditions etc. to constrain the distortion tensor. The full-pattern projective matching approach automatically treats the crystallographic correlations between all the Kikuchi bands and zone axes, as compared to multi-ROI approaches in which the internal consistency of changes in these features is not necessarily included from the outset.

\textit{Experimental Details.}
For the  EBSD investigation presented in this paper, we have used a Versa 3D (FEI) field-emission scanning electron microscope with a Hikari detector (EDAX) at $120\times 120$ pixel resolution, 20\,kV primary beam voltage, a beam current of 15\,nA, and a measurement time of 20\,ms per pattern. The raw EBSD patterns were stored for the subsequent analysis using the pattern matching approach.

The best fit pattern parameters were determined by the optimization of the normalized cross-correlation coefficient (NCC) \cite{gonzalez2007DIP3} $r$ ($0<|r|<1$) between the simulated data and the experimental Kikuchi patterns.  
We used the Nelder-Mead simplex method \cite{neldermead1965tcj} to find the local maximum of the NCC between experiment and simulations, with start parameters at an orientation obtained from the manufacturer EBSD software. 
For the dynamical electron diffraction simulations we have used the Bloch wave approach according to \cite{winkelmann2007um}. 
In the NCC optimization procedure, the test Kikuchi patterns are reprojected from a stored global pattern according to the current values of the projection parameters, the orientation, and the tetragonal lattice distortion, then the NCC is calculated, and updated projection and distortion parameters are chosen for the next iteration according to the simplex approach until convergence to a local maximum with changes of the NCC $\Delta r < 10^{-4}$ is achieved. 
The austenite phase in the map area has been used for calibration of the detection geometry via pattern matching to dynamical simulations, assuming a cubic fcc structure \cite{nolze2018amat}.
EBSD pattern processing involved high-pass filtering to remove the slowly-varying background and normalization of the Kikuchi diffraction patterns to a mean value of $\mu=0.0$ and standard deviation $\sigma=1.0$ for visualization.

\begin{figure}[htbp]  %
	\includegraphics[width=16cm]{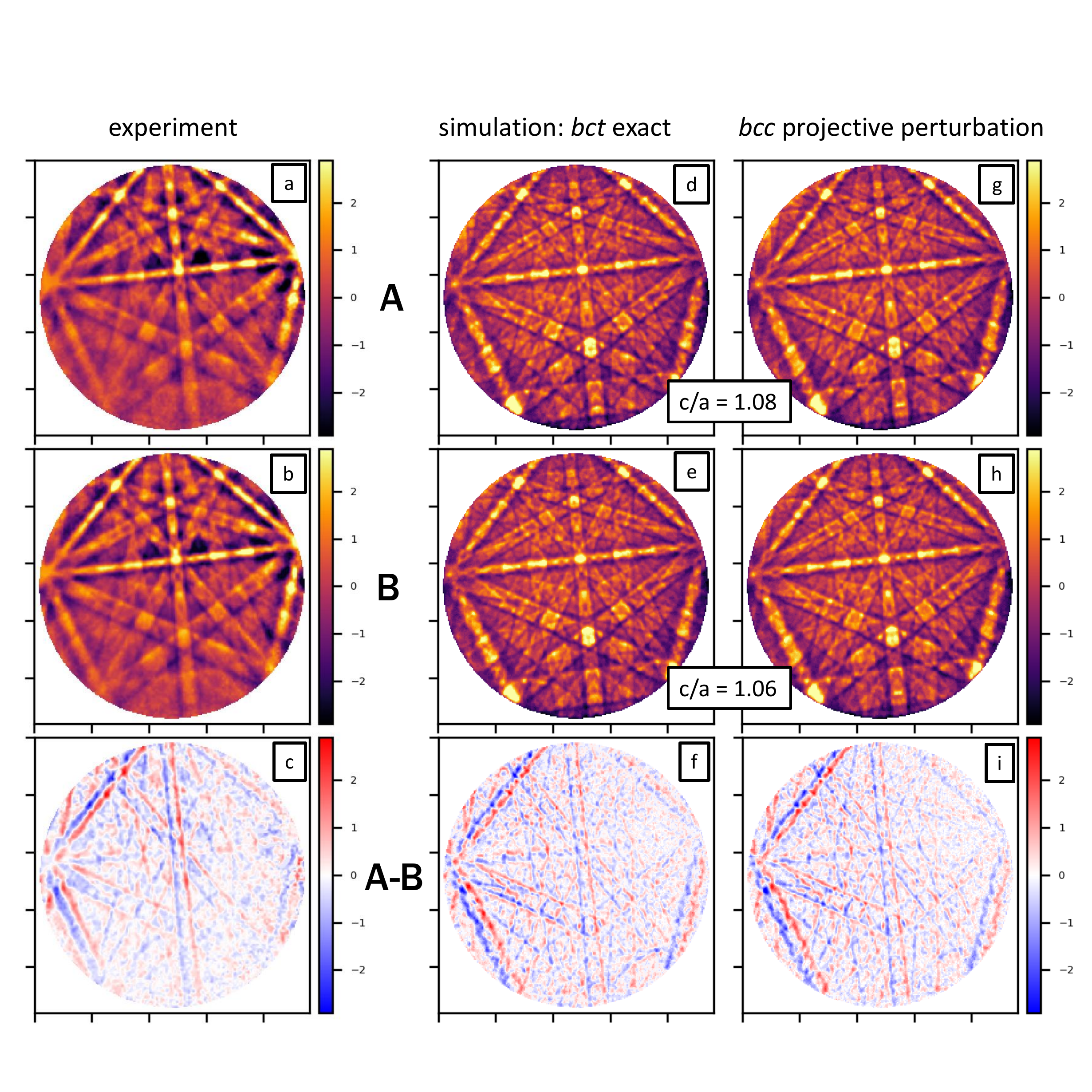}
	\caption{Effect of a locally changing tetragonality on Kikuchi patterns from martensite.\\
	\textbf{Left column}: Experimental Kikuchi patterns (a,b) from positions A and B in the map shown in Fig.\,\ref{fig:sem}, (c) experimental pattern difference.\\
	\textbf{Middle column}: simulated exact Kikuchi patterns for Fe bct structures with fixed c/a ratios,\\
	\textbf{Right column}: best fit Kikuchi pattern approximations for a continuously varying c/a ratio obtained by projective transformation of simulated bcc reference data.}
	\label{fig:patterns}
\end{figure}

\textit{Results.}
In Fig.\,\ref{fig:patterns}, we illustrate the effects of a change in local tetragonality on the observed Kikuchi patterns from locations A and B in the SEM image of Fig.\,\ref{fig:sem}.
In Fig.\,\ref{fig:patterns}(c) we show the difference of the experimental patterns in Fig.\,\ref{fig:patterns}(a) and (b). The red and blue regions in the left half of the difference pattern show where the largest changes occur, i.e. in the region around the \hkl[111] zone axis of the bct lattice (see Fig.\,\ref{fig:bct_lattice}).  

The middle and the right column of Fig.\,\ref{fig:patterns} allow a comparison of the experimental patterns to simulations. 
In the middle column, we show the result of exact dynamical diffraction simulations for two bct structures with $c/a$-ratios of (d) 1.08 ($r=0.651$) and (e) 1.06 ($r=0.661$), respectively. These patterns have been obtained after testing against simulations for $c/a$-ratios in the range of 1.0 to 1.1, with a 0.01 step size.  
For comparison, in the right column of Fig.\,\ref{fig:patterns}, we show the result of a variable projective transformation of simulated global pattern data for a bcc structure ($c/a=1.0$), where allowed a scaling transformation along the the c-axis of the initial bcc crystal coordinate system.
After testing for the best of 3 possible pseudosymmetric orientation solutions, we obtained best fit $c/a$ values of (g) 1.084 ($r=0.646$) and (h) 1.062 ($r=0.660$).
The close agreement of the NCC $r$-values for the exact and the approximated pattern indicates that the projective approximation captures the main features of the experiment in both cases equally well. 
Because the $r$-values for the approximate pattern will decrease with increasing distortion, we can use the difference of $r$ relative to the reference for an assessment of the quality of the approximation. 
In the bottom row of Fig.\,\ref{fig:patterns}, we see that the experimental difference pattern A-B is well reproduced by both of the analysis approaches, with the variable projective transformation not requiring any additional dynamical simulation beyond the bcc reference structure.
From the exact results shown in the middle column (d,e), it is also obvious that a refinement of the variably transformed bcc patterns shown in Fig.\,\ref{fig:patterns}(g,h) is possible via re-fitting to new \textit{bct} reference structures with an updated $c/a$ ratio near the values obtained in the initial fit step to the distorted \textit{bcc} start structure.

\begin{figure}[tb!]   %
 \centering
 \includegraphics[width=12cm]{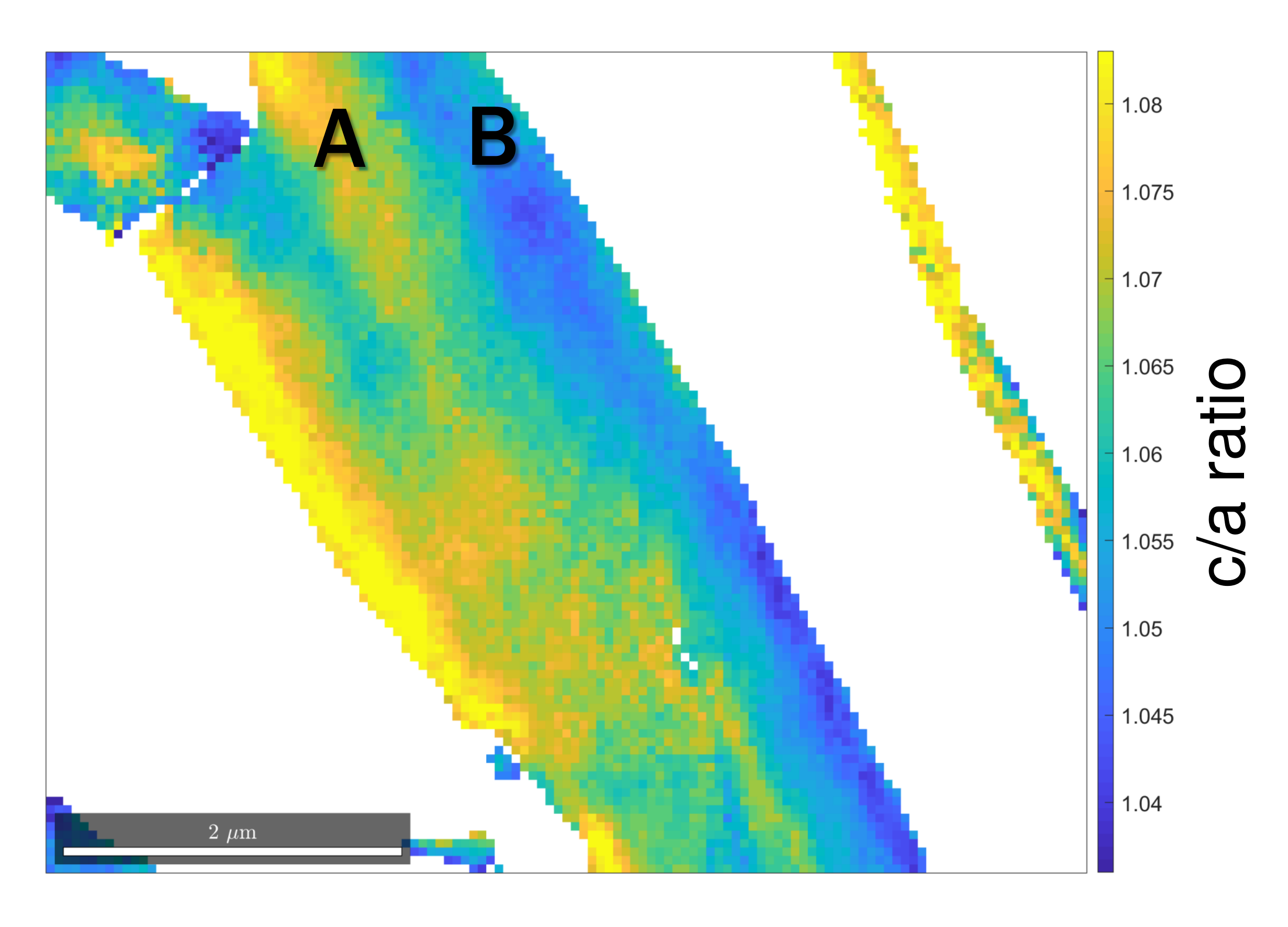}
	\caption{ Local c/a ratio in the Martensite, obtained by variable projective pattern matching}
	\label{fig:tetmap}
\end{figure}

Finally, applying the projective pattern matching to the full map of  Kikuchi patterns measured from the martensite grains, we obtain a map of the local $c/a$-ratio which is shown in Fig.\,\ref{fig:tetmap}(a). We can observe a range of tetragonalities from about 0.03 to 0.08, which, for this specific grain, is consistent with the averaged value of 0.07 estimated in the X-ray measurements. 
We expect a more detailed, statistically significant picture by mapping the tetragonal distortions in larger sample areas. This will allow better insights into the details of the local martensite tetragonality formation, especially concerning the driving forces and stresses which are involved in the interplay between local carbon content and lattice distortion.

\textit{Error Assessment.}
For a reliability analysis of our EBSD pattern matching approach, we have used Monte-Carlo simulations of the EBSP formation, detection and analysis pipeline acting on simulated raw EBSD patterns with known, varying parameters for orientation, tetragonality and projection geometry. 
In addition to the dynamical electron diffraction effects of the Kikuchi pattern, we have taken into account the various noise sources, and optical effects \cite{britton2010um,hansen2017mm}, as well as phenomenological models for the EBSD background signal and the excess-deficiency effect on the Kikuchi bands.
These simulations give us an estimation of the frequency distribution of the NCC $r$-values for the simulated experimental conditions and thus allow us to estimate confidence intervals for the fitted parameters in the pattern matching approach.

The limited number of 8 free parameters which are combined in a non-trivial way in the elements of the $(3x3)$ matrix $\mathbf{H}$ \cite{alkorta2017um} shows that an accurate knowledge of the projection center is necessary to calibrate the coordinate transformations in the detector plane, which is a key experimental problem in HR-EBSD investigations.
The estimated tetragonality will depend systematically on the correct calibration of the projection center, for which typical error values of about 0.2--0.5\% of the pattern width have been estimated in the literature \cite{maurice2011um,mingard2011um}.
Taking a value of $0.5\%$ as a conservative upper-limit estimate, we find that a systematic shift in tetragonality of about 0.01 will be induced by this error in PC for the main observed grain in the current study. The actual PC we used for the current map was determined from the fcc austenite region which was assumed as perfectly cubic.

Assuming a fixed PC and comparing simulated patterns for a known tetragonality in the range of 0.0 to 0.1, we find for the observed orientation of the Martensite grain a systematic underestimation of the tetragonality by 0.005, within a standard deviation of 0.001. 
These values are consistent with the noise which we observe in the tetragonality map of Fig.\,\ref{fig:tetmap}.

Our error assessment suggests that we achieve a precision in the order of $>10^{-3}$ for the tetragonality at a modest pattern resolution of $100^2$ pixels in the current study, compared to typically about $1000^2$ pixels for typical HR-EBSD work.
As the projective approximation which we used here becomes better at small distortions, the method presented here can be combined with iterative best-fit methods which use reference patterns at varying levels of distortion as discussed e.g. in \cite{jackson2016mm,tanaka2018mm} to improve the precision of the results. 

While the observed tetragonalities in the current application stay below 0.10, we note that the distortion approach is flexible enough to accomodate the full range of the hypothetical $c/a$-ratios in the Fe $\gamma - \alpha $ (fcc-bcc) transition, as shown above in Fig.\,\ref{fig:bct_lattice}. Quantitatively, this means that for the $fcc$ Kikuchi pattern in position C of the map we can find a best fit value for the $c/a$-ratio of 1.405 (near the ideal $\sqrt{2}$ for fcc) when we consider pattern C as resulting from the distorted bcc reference data ($c/a=1.0$). The  best-fit NCC $r$-value of the distorted bcc pattern is $r=0.53$, which still compares well to $r=0.67$ for the fit to the simulated exact Fe fcc structure. The close relationship between the atomic positions in the $\alpha$-Fe and $\gamma$-Fe crystal structures, irrespective of their description by specific crystallographic symmetry groups, leads to a sufficiently clear correspondence between the distorted lattice plane traces and their intersections (which are transformed exactly) and the related Kikuchi bands (which are transformed only approximately). 

In summary, we have presented a crystallographic approximation for the determination of variable lattice distortions using full-pattern simulations of Kikuchi diffraction in the SEM. This improves the options to extract approximate lattice parameter information even from conventional, routine EBSD data.
Some relevant materials which we can envision include technologically and geologically important structure types, like, for example locally deformed materials in metallurgy, distorted thin films for semiconductor applications, or the cubic-tetragonal-orthorhombic-rhombohedral line of related structures of perovskites under varying chemical and environmental conditions.
In this way, we expect that the approach presented here will be useful for the analysis of various important structure types and their relationships by microcrystallographic methods in the SEM.



%

\end{document}